\documentclass[conference]{IEEEtran}

\usepackage{amsmath}
\usepackage{graphicx}
\usepackage{epstopdf}
\usepackage{colordvi}
\usepackage{url}

\newcommand{\beq}{\begin{equation}}
\newcommand{\eeq}{\end{equation}}

\begin{document}
%
\title{Switching Times in Fabry-Perot Measurements}

\author{
\IEEEauthorblockN{Paolo Addesso}
\IEEEauthorblockA{\small
Dept. of Computer Science, Electrical Engineering\\ and Applied Mathematics\\ 
 University of Salerno\\
Via Giovanni Paolo II, 132, 84084 Fisciano, IT \\
 Email: paddesso@unisa.it
}
\and
\IEEEauthorblockN{Vincenzo Pierro}
\IEEEauthorblockA{\small 
Dept. of Engineering\\
University of Sannio\\
C.so Garibaldi 107, 
82100 Benevento, IT\\
Email: pierro@unisannio.it}
\and
\IEEEauthorblockN{Giovanni Filatrella}
\IEEEauthorblockA{\small
Dept. of Science and Technology \\
 University of Sannio \\ Via Port'Arsa 11, 82100 Benevento, IT \\
 Email: filatrella@unisannio.it \\ 
}
}

\maketitle

\begin{abstract}
We show how to analyze the motion of very low dissipation suspended mirrors in a Fabry-Perot. The very precise measurements of the mirrors motion can be determined, also in the presence of a disturbing noise, by means of the sudden reflectivity changes in special points of the  mirrors positions.
When the mirrors cross such positions, the effective opto-mechanical potential that arises in the device is (roughly) at a maximum. We show that the motion cross such potential maxima is not only confused by the presence of noise, but also favoured by noise itself that induces hoppings. Thus, the measurements of the times at which the crossings occur can be exploited to identify the properties of the applied signal. We also show how to circumvent the difficulty of the extremely long transient that occur in the system analyzing the escape average time with two different methods: a direct sample average and the indirect estimate from the tail distribution. Numerical simulations and physical insight suggest that the indirect estimate, through the analysis of the distribution tails with an appropriated cut off is robust against the disturbances that arise from the presence of transient dynamics. 
\end{abstract}
\IEEEpeerreviewmaketitle

\section{Introduction}
Astronomical events can be detected to validate and discriminate general relativity theories \cite{Deruelle84,Drever83}. Furthermore, the detection of gravitational radiation might open a new observational window on the Universe \cite{InterfeWeb}. Such astronomical relevant  quantities are investigated with Michelson-Morley interferometers, where the arms are made of Fabry-Perot (FP) containing very low dissipation suspended mirrors \cite{Rakhmanov98}. 
The measurements ultimately consist of accurately detecting the mirrors position change. 
The position fluctuations are a fraction of the light wavelength measured over the macroscopic (up to kilometers) distance between the mirrors. 
Obviously, noise is a major obstacle, for the motion of the mechanical part is disturbed by external seismic and internal thermal random sources that blurry the signals.
One of the most important noise sources is thermal noise in the high-reflectivity dielectric coatings of the test masses.
Reduction of  such coating thermal noise is crucial to reach, or hopefully surpass, the standard quantum-noise level \cite{Principe15}.
The motion of the mirrors amounts to oscillations in a multistable opto-mechanical potential \cite{Aguirregabiria87,Pierro94}. 
To test the electromagnetic mirrors performances  and the noise effective temperature \cite{Villar10,Bodiya12}, we propose to measure the escape time from a minimum, instead of the full trajectory. 
In analogy with research in Josephson junctions dynamics \cite{Marchesoni97} where escapes have been employed to characterize Poissonian \cite{Pekola04} or non Gaussian noise sources \cite{Valenti14}, escape times can be also used to characterize thermal noise intensity on pendular Fabry-Perot \cite{Addesso15}. 
In this work we propose to extend the application of escape times analysis to the case where a signal is embedded in noise.
Escape times are easier to measure, while their information content is not much depleted respect to the full trajectory \cite{Addesso13}. 
The nonequilibrium character of the Fabry-Perot system poses an intrinsic difficulty, inasmuch accurate measurements require two contradictory features: low dissipation to reduce thermal noise, and transients suppression to measure statistical properties. 
We show that the estimate of the mean exit time from the distribution of the escapes can be performed with two techniques: sample mean estimator and maximum likelihood analysis. Comparing the two methods with numerical simulations of the electromechanical system, we find that the estimate through the tails of the distribution is less sensitive to transient dynamics. 

\begin{figure}[t]
\centering
\centerline{(a)}
\includegraphics[width=3in]{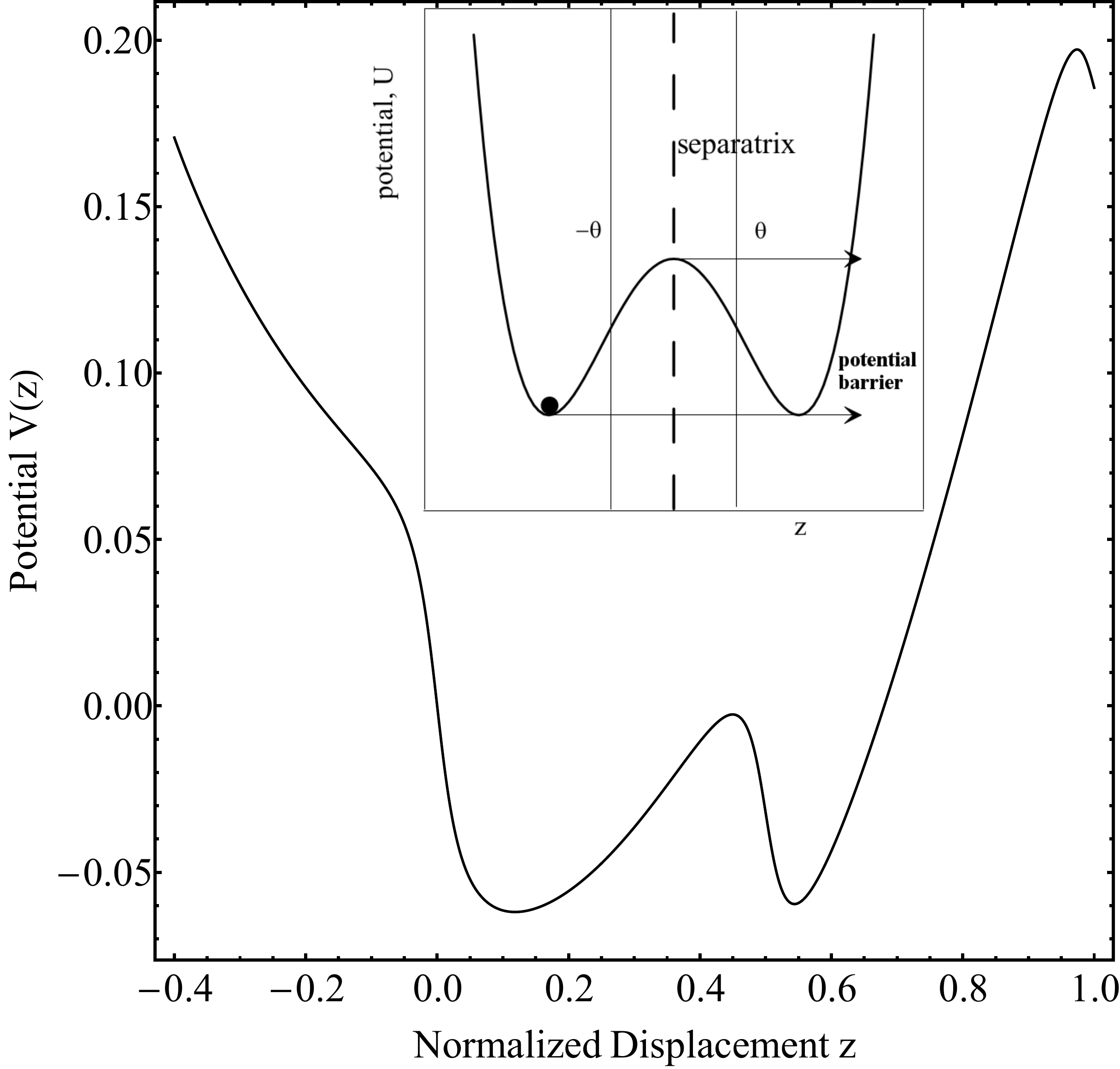}
\vspace{0.5cm}\phantom{$d=0$}
\centerline{(b)}
\includegraphics[width=2.5in]{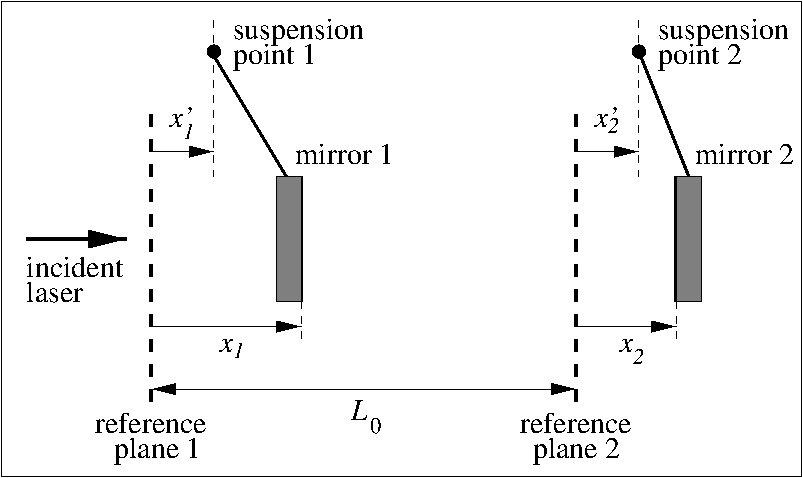}
\caption{(a) The potential of the mirrors . Compared to the standard quartic potential. (b) Schematic of a Michelson-Morley arm making the FP interferometer.
}
\label{fig:potential}
\end{figure}

\section{Mathematical model}
\label{sect:model}

The mirrors depicted in Fig. \ref{fig:potential}b are  subject to the gravitational torque and the radiation pressure. The latter is periodic with a spatial period that depends upon the radiation wavelength. Together with the ordinary pendulums potential it gives rise to the following potential:

Equation (\ref{eq:zeta_norm}) describes the motion of a particle in the potential of Fig.\ref{fig:potential}a.
\begin{eqnarray}
V(z)= & - & \frac{\Pi_M (
	\arctan (\sqrt{{\cal F}+1} \tan(2\pi  z)
	)+ \pi \lfloor 2 z+\frac{1}{2}\rfloor )}
{2 \pi 
   \sqrt{{\cal F}+1}}   \nonumber \\
& + & \frac{1}{2} z^2    -z A  \sin \left(\omega t \right),
   \label{eq:potential_FP}
\end{eqnarray}

\noindent where $\lfloor \cdot \rfloor$ denote the floor truncation function.
Here $z$ is interpreted as the displacement of the terminal mass normalized respect to $\lambda$, the wave length of the incident laser radiation. 
The mirrors separation dynamics $z$ in the potential of Eq.(\ref{eq:potential_FP}), see Fig. \ref{fig:potential}, is governed by the following differential equation:
\beq
\frac{d^2z}{dt^2} + \gamma \frac{dz}{dt} = - z + \Pi_{M}{\cal A}(2\pi z) 
+ A \sin\left( \omega t \right) +  \xi(t),
\label{eq:zeta_norm}
\eeq
\beq
< \xi(t),\xi(t')> = 2D\delta(t-t').
\label{eq:correlator}
\eeq
\noindent

\noindent 
In Eqs.(\ref{eq:zeta_norm},\ref{eq:correlator}), the symbol $\xi(t)$ denotes  the random process incorporating the whole noise that affects the system masses.
Furthermore,  $\Pi_{M} = R_{M}/\mu \lambda \omega_0^2$ is the normalized radiation pressure coefficient, $\mu$ and $\omega_0$ are the reduced mass of the pendulum and the mechanical frequency of the end mass, respectively. 
Time is normalized respect to the inverse of the natural frequency, $\omega_0$.
Actual systems achieve such small damping as $\gamma = 10^{-6}$\cite{Drever83}.
This extremely low damping makes the simulations relatively difficult \cite{Sivak13}, but also simplifies the analysis that can be performed in the limit of very small friction\cite{Mannella04}. The normalized input power is $R_{M}= 2 P_{M}/c$ where $P_M$ is the laser input power and $c$ is the vacuum speed of light.
The friction constant $\gamma=\tilde{\gamma}/\omega_0$ is given by the pendulum dissipation constant $\tilde{\gamma}$ divided by $\omega_0$ . 
The function ${\cal A}$ is the Airy function ${\cal A} =1/(1+{\cal F} \sin^2{\phi})$, where 
${\cal F}=\pi\sqrt{r_1 r_2}/(1-r_1r_2)$ is the Finesse of the cavity ($r_{i}$ and $t_i$ are the reflection and transmission coefficients, respectively; $i=1,2$ refers to the mirrors, see Fig. \ref{fig:potential}), and  $\phi=2\pi z$ is the phase of the circulating light.
The phase $\phi$ also determines the half-width $\delta \phi$ of the resonance, $\delta \phi = \pi/2 {\cal F}$. 
The maximum stored power corresponds to the peaks of the Airy function ($\phi=n\pi$).

We have assumed in Eq. (\ref{eq:correlator}) that the noise intensity $D$ is dominated by the external sources, and therefore does not obey the dissipation fluctuation theorem \cite{Risken89}. Therefore, the Kramers escape time $\tau_K$ \cite{Kramers40} is modified as follows:
\beq
 \tau_K  = \tau_0 \exp\left( \frac{\gamma \Delta V}{D}\right).
\label{kramers}
\eeq
In Eq. (\ref{kramers}) the activation energy $\Delta V$ is an effective barrier that incorporates the periodic oscillations induced by the term $A \sin \left( \omega t \right)$. 
We underline that the oscillating term is central  in the gravitational measurements, for the measurements amount to infer the presence (and the properties) of the sinusoidal term via the analysis of the escape times.
The escapes are convenient for measurements, in that when the mirrors cross the points of sharp change of reflectivity, a sudden increase of the reflected power occurs, as the injected power in between the mirrors hits a maximum.
Thus, the time elapsed to overcome the barrier, the escape times (or the first passage times \cite{Risken89}), given by Eq.(\ref{kramers}), can be detected without mechanical contacts. 

To analyze the escape times, it is convenient to approximate the potential (\ref{eq:potential_FP}) with the more standard 
\beq
\frac{d^2 z}{dt^2} + \gamma \frac{dz}{dt} =  a z - b z^3 + A \sin\left( \omega t \right) +  \xi(t),
\label{eq:quartic}
\eeq
\noindent 
that is characterized by the deeply investigated quartic potential:
\beq
U(z)= \frac{a z^4}{4} - \frac{b z^2}{2} - z A  \sin \left(\omega t \right)
   \label{eq:potential_quartic}
\eeq
The physical interpretation of the coefficients $a$ and $b$ in Eq.(\ref{eq:quartic}) is that the minima occur at $x=\pm \sqrt{{b}/{a}}$ and the activation energy reads $ \Delta U = {b^2}/{4a}$. Thus, one can calibrate the coefficients to reproduce the main physical properties of the FP potential, see Fig. \ref{fig:potential}(a). 

By way of conclusion of this part, we would like to emphasize that the present analysis might prove useful in a larger context, as the escape times are not only convenient (as in the present case), but sometimes are the only available quantity (e.g., Josephson junctions \cite{Addesso12}, climate changes \cite{Benzi82}, and quantum measurements \cite{Aspelmeyer14}). 

\begin{figure}[t]
\centerline{(a) }
\centerline{\includegraphics[scale=0.28]{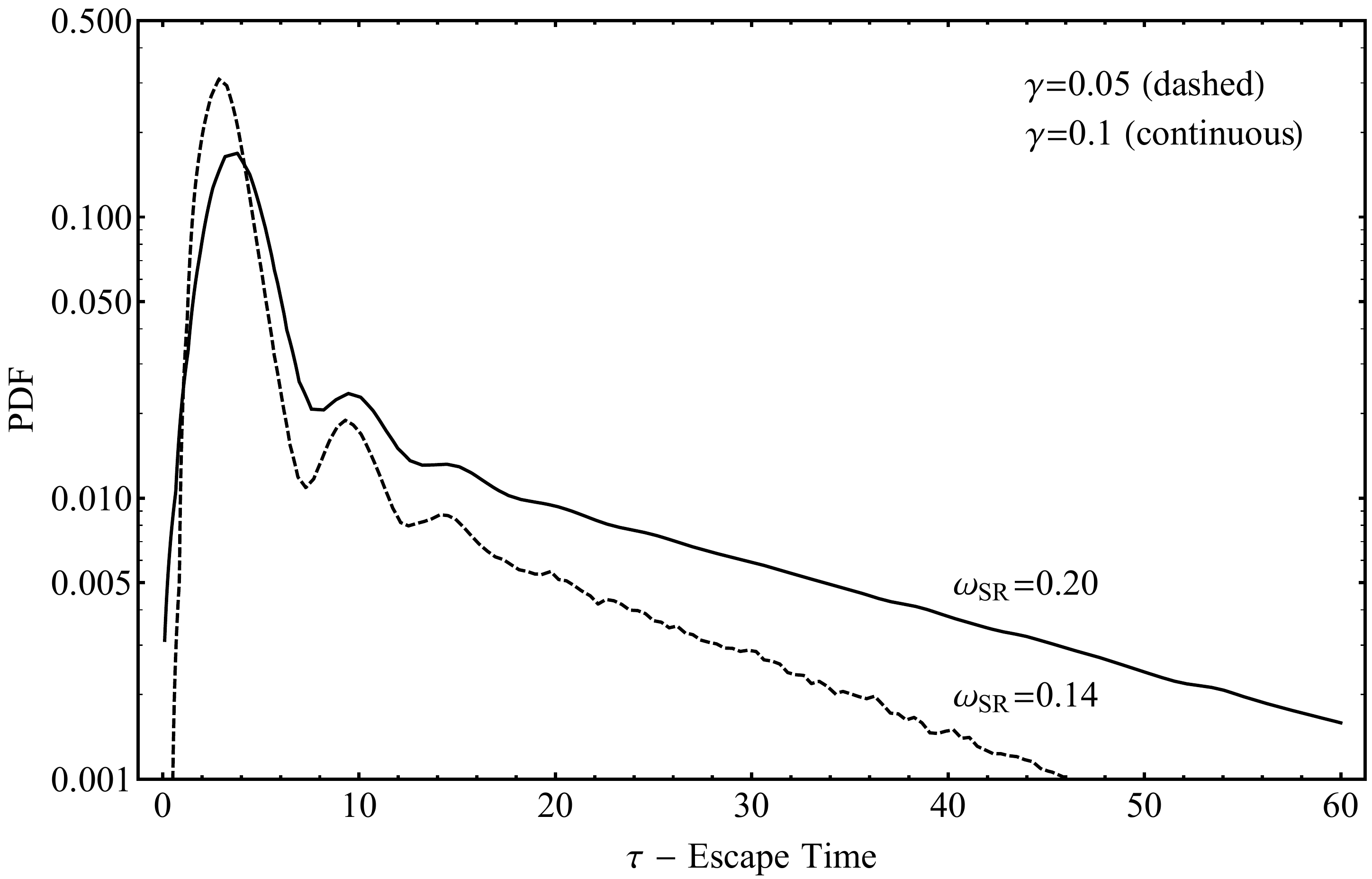}}
\centerline{(b) }
\centerline{\includegraphics[scale=0.28]{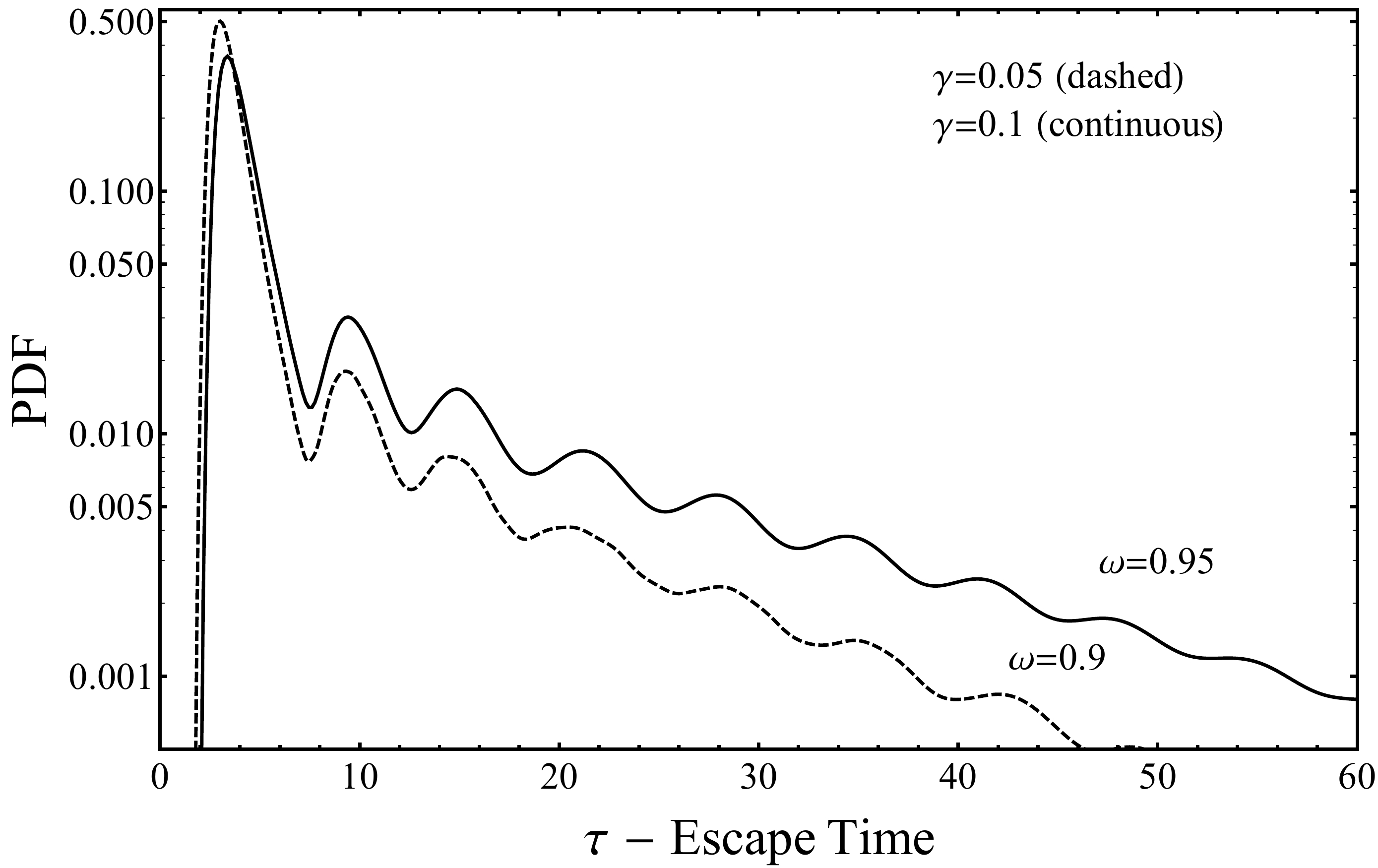}}
\centerline{(c)}
\centerline{\includegraphics[scale=0.25]{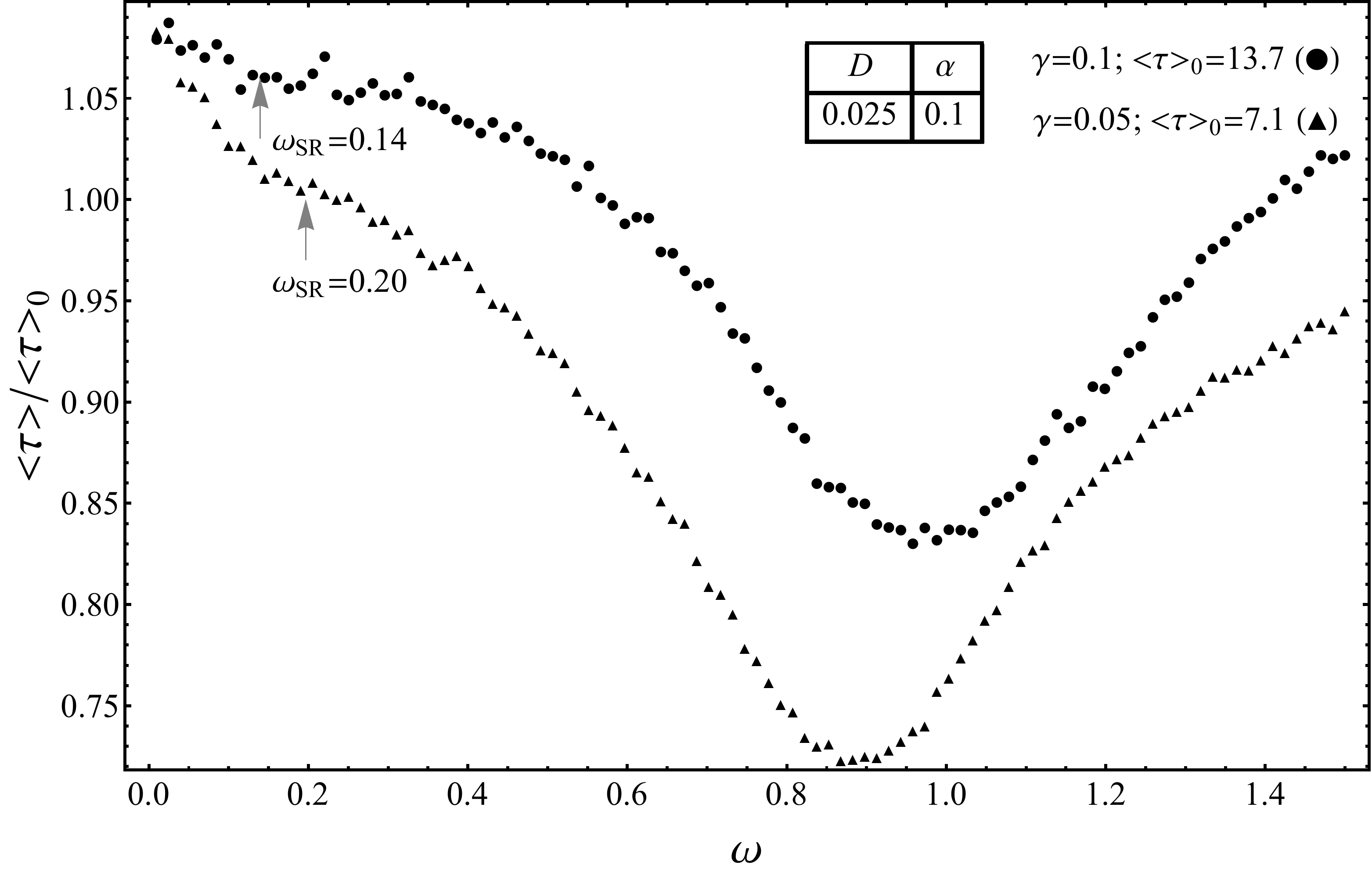}}
\caption{
(a) Probability Density Function (PDF) of the escape times from simulations of Eq.(\ref{eq:quartic}) with $A=0$. The expected stochastic resonance frequency, on the basis of the tail slope, is denoted by $\omega_{SR}$, see Eq.(\ref{SRfreq}).
(b) Probability Density Function (PDF) of the escape times from simulations of Eq.(\ref{eq:quartic}) with $A\neq 0$ and frequency $\omega = 0.9$ and $\omega = 0.95$ for the dotted  and solid curves, respectively. 
The panel shows the sensitivity of the escape time distribution to the presence of a coherent external sinusoidal signal. 
(c) Sensitivity to the external frequency of the averaged escape time obtained from distributions displayed in (b) . The arrows indicate a region where the SR occurs \cite{Addesso12,Pierro04,Pierro98,Gammaitoni98}.
}
\label{fig:distribution}
\end{figure}

\section{Measuring the average escape times}
\label{sect:estimate}
It is very difficult to derive analytically the escape time Probability Distribution Function (PDF) in underdamped systems and in the presence of a sinusoidal drive.
This is witnessed by the intense research to retrieve analytical approximations in several limiting cases \cite{Zhou90,Berglund05}.
An approximated form of the PDF for a time independent potential reads:
\beq 
f(\tau) \approx \exp \left[-  \frac{ \tau}{\langle \tau \rangle} \right]
\label{eq:distribution}
\eeq

\noindent where $\langle \tau \rangle $ is the average escape time. 
The approximated nature of Eq.(\ref{eq:distribution}) is evident in Fig. \ref{fig:distribution}(a); in particular it is clear that deviations occur at low values of the escape time $\tau$, while the exponential behavior is found in the asymptotic regime. 
This is due to the presence of resonant frequencies associated to the minima and the maxima of the potential (see Fig. \ref{fig:potential}a) that affects the short transients.

In Fig. \ref{fig:distribution}(b) we display the PDF of the escape times in the presence of an external  drive, $A\neq 0$ in Eq.(\ref{eq:quartic}). 
We denote with $\langle \tau \rangle $ the average escape time, while the average escape time without forcing ($A=0$) is denoted by $\langle \tau \rangle_0 $. 

As noted earlier, the approximated distribution Eq.(\ref{eq:distribution}) is not the actual distribution, see Fig. \ref{fig:distribution}(a), for a number of reasons that we shall discuss in the following. 
However, an important feature that is evident in this figure is the dependence of the distribution upon the drive angular velocity $\omega$. 
In particular, there is a specific value:
\beq
\omega_{SR}\simeq \pi / \langle \tau \rangle _0
\label{SRfreq}
\eeq

\noindent  where the distribution exhibits the so called Stochastic Resonance (SR) phenomenon \cite{Gammaitoni98}: the hopping is favoured by the coincidence of two time scales, i.e. when the drive period and the noise assisted escape match each other. 
In fact, as we show in Fig. \ref{fig:distribution}(c), the average escape time shows an inflection point in the neighbours of the SR values given by Eq.(\ref{SRfreq}). 
Shortly, some physical properties of the sinusoidal drive applied to the system can be deduced from the parameter dependence of  $\langle \tau \rangle$. 

\begin{figure}
\centerline{\includegraphics[scale=0.4]{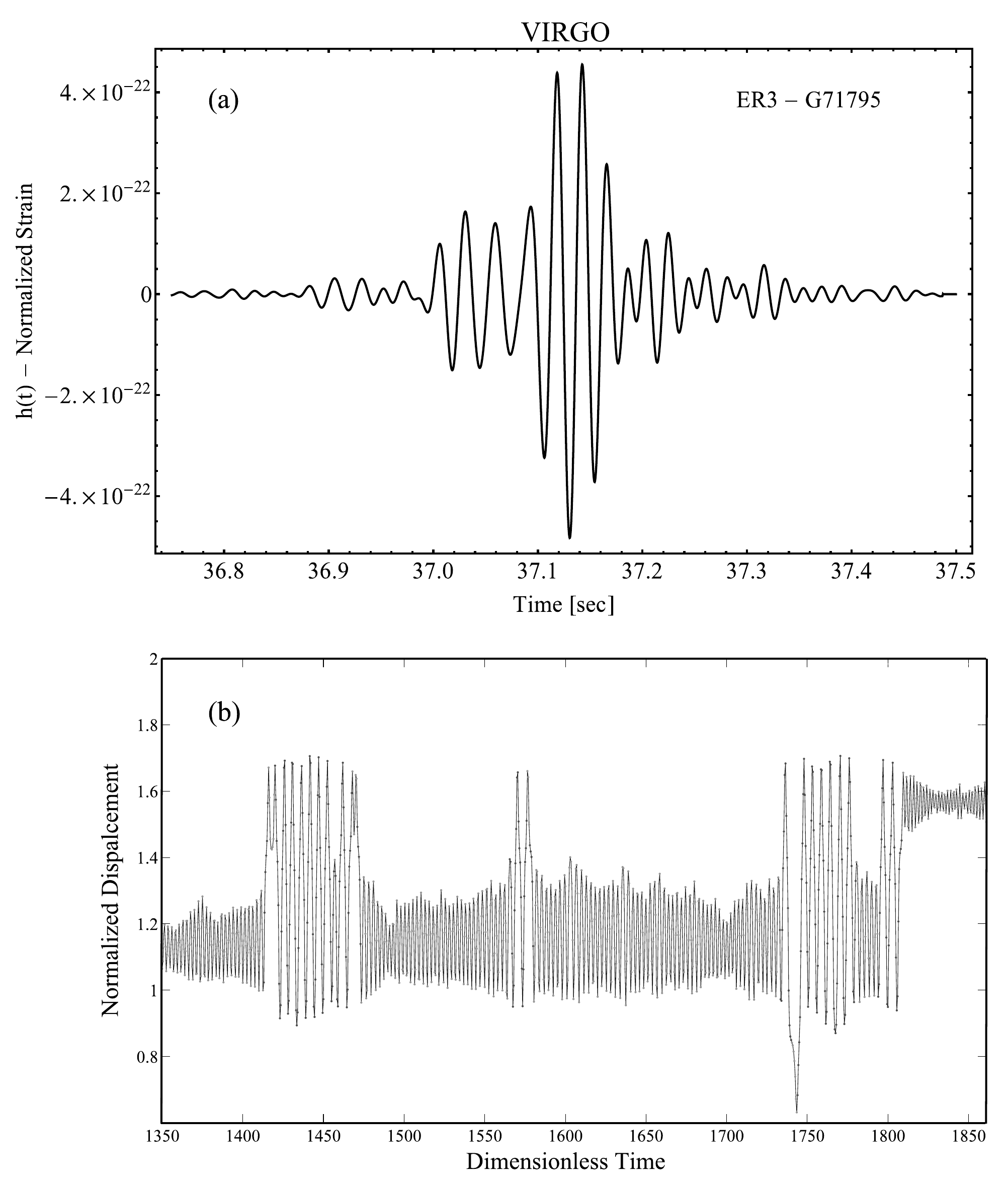}}
\caption{
(a) a typical transient measured in Virgo interferometers.
(b) a simulated transient in the system governed by Eq.(\ref{eq:quartic},\ref{eq:potential_quartic}). 
The parameters of the simulations are $~{\cal F}=1000$, $D = 0.0005$, $\gamma = 10^{-2}$ and $\Pi_M=2$.
The similarity of the  behaviours in (a) and (b) suggests that in underdamped multi-stable systems, noise itself generates transitions among the nearly metastable states, that produce large oscillations.
}
\label{fig:transient}
\end{figure}

Coming back to the actual escape time distribution of Fig. \ref{fig:distribution}(a), we note that unfortunately the low part of the distribution, corresponding to the short escape times, is affected by two main problems: the inertial effects that prevent an instantaneous passage across the barrier and spurious transient oscillations over the unstable equilibrium at the maximum. Moreover, the escape times are disturbed by the long transients \cite{Rampone13} entailed by the low friction of the system, see Fig. \ref{fig:transient}.

We first discuss how to deal with the high frequency spurious oscillations. We propose to employ a threshold $\theta$ to decide if the system has effectively passed the barrier maximum. 
In fact, a difficulty arises in determining the passage across a boundary with simulations of  stochastic differential equations \cite{Mannella01}. 
To overcome this difficulty, we propose to use an effective threshold value $\pm \theta$,  see the inset of Fig. \ref{fig:potential}(a), to decide if the system has effectively passed the separatrix. 
If the particle is initially in the left hand side of the potential, ($x<0$), one only counts a passage across the threshold $+\theta$, beyond the separatrix in the descending part of the potential. 
The same is true for the reverse process: if the system is initially in the right hand side of the potential ($x>0$), an escape is defined as the passage across $-\theta$.
The escape time is analogous to the first passage time  (in the stationary state of particle) to reach the threshold $\theta$ starting from the position $x<0$.
To determine a suitable value for the parameter $\theta$, we first observe that the distribution for long escape values is asymptotically  exponential (characteristic of the absence of the applied external drive), the distribution given by Eq.(\ref{eq:distribution}), while for low escape time it deviates from the exponential distribution.
In accordance with these observations, we chose the appropriated $\theta$ value imposing that $\omega_{SR}$  agrees with the asymptotic exponential decay.
This is done assuming an exponential asymptotic behavior similar to Eq.(\ref{eq:distribution}), but with another  parameter, $\tau_e$ (not necessarily coinciding with the average escape time $\langle \tau \rangle $):
\beq
f(\tau) = {\mathcal N } \exp \left( \frac{\tau}{\tau_e}\right) \,\,\, {\rm for} \,\,\,  \tau > \tau_c.
\label{eq:fitting}
\eeq
Thus, $\tau_c $ is a cut off time, ${\mathcal N}$ is a normalization factor. The cutoff arises because of physical reasons: the transient relaxations \cite{Lanzara97} around the metastable minima and  the fast oscillations associated to the interwell dynamics. Both effects are exacerbated by the low damping. 

\begin{figure}
\centerline{\includegraphics[scale=0.36]{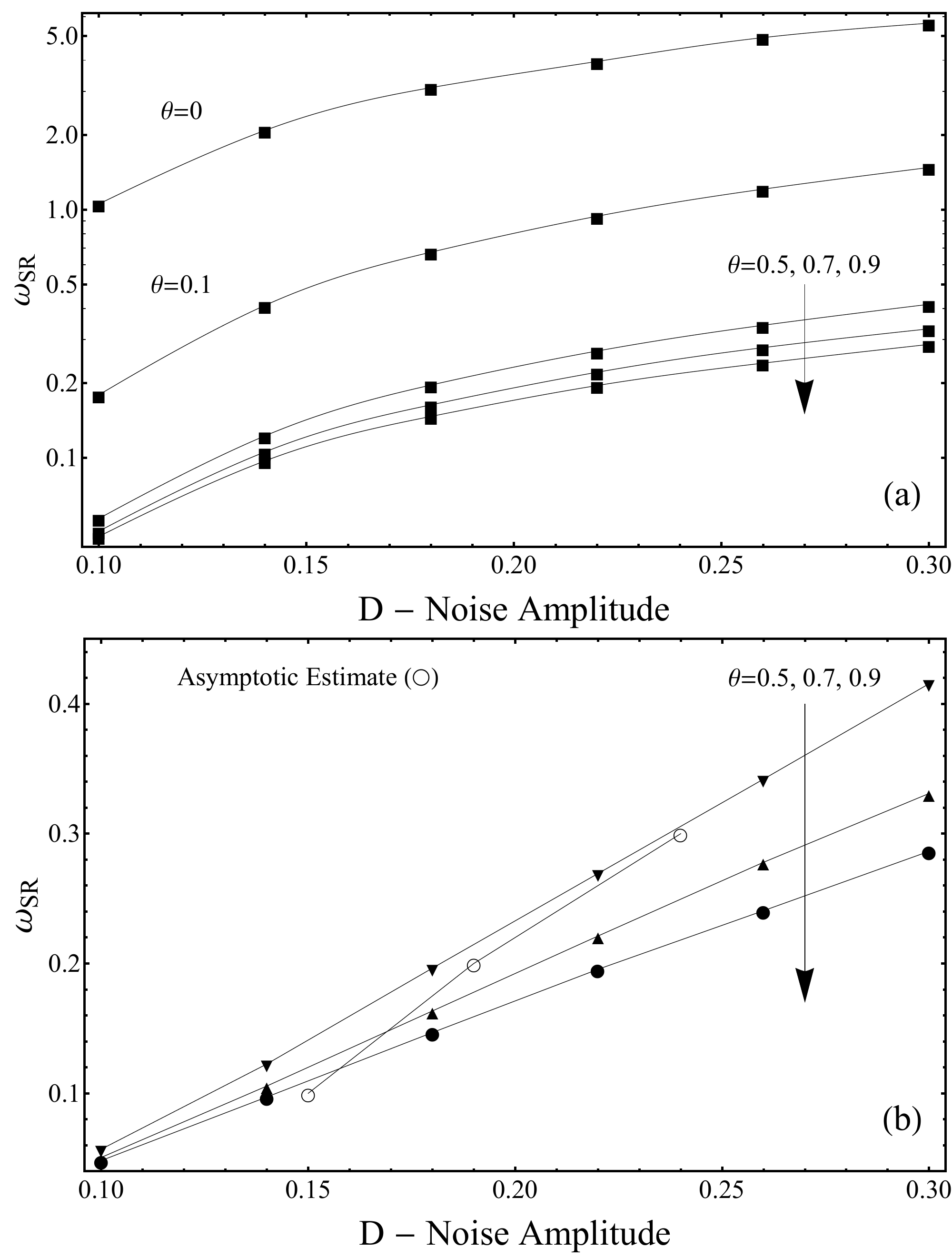}}
\caption{
(a) Analysis of the threshold dependence of the SR frequency estimate as a function of noise. 
(b) close-up of part (a) where circles denote the sample mean estimate obtained in the absence of the external drive.}
\label{fig:threshold}
\end{figure}

We display in Fig. \ref{fig:threshold} a typical estimate of the stochastic resonance frequency through Eq. (\ref{SRfreq}). The collection of escape times depends upon the choice of the threshold $\theta$ to discriminate a passage from a basin to the other, and so does the average in the denominator of (\ref{SRfreq}). Consequently,  a good choice for the $\theta $ parameter entails a stabilization of the estimated $\omega_{SR}$ at different values of the noise intensity $D$.
Fig.\ref{fig:threshold} demonstrates that for $\theta >0.5$ a stable $\omega_{SR}$
estimated can be achieved. This value corresponds to the inflection points of the potential, see Fig. \ref{fig:potential}a. 
A reasonable agreement is found for $\theta \simeq 0.6$, that is the value employed in this paper. The value is also supported by comparison with the results of the averaged escape time in the absence of external drive -- see Fig. \ref{fig:threshold}b. This is important, for in the case of a pure noise term, without external drive, spurious oscillations are likely to be depressed.

\begin{figure}
\centerline{\includegraphics[scale=0.28]{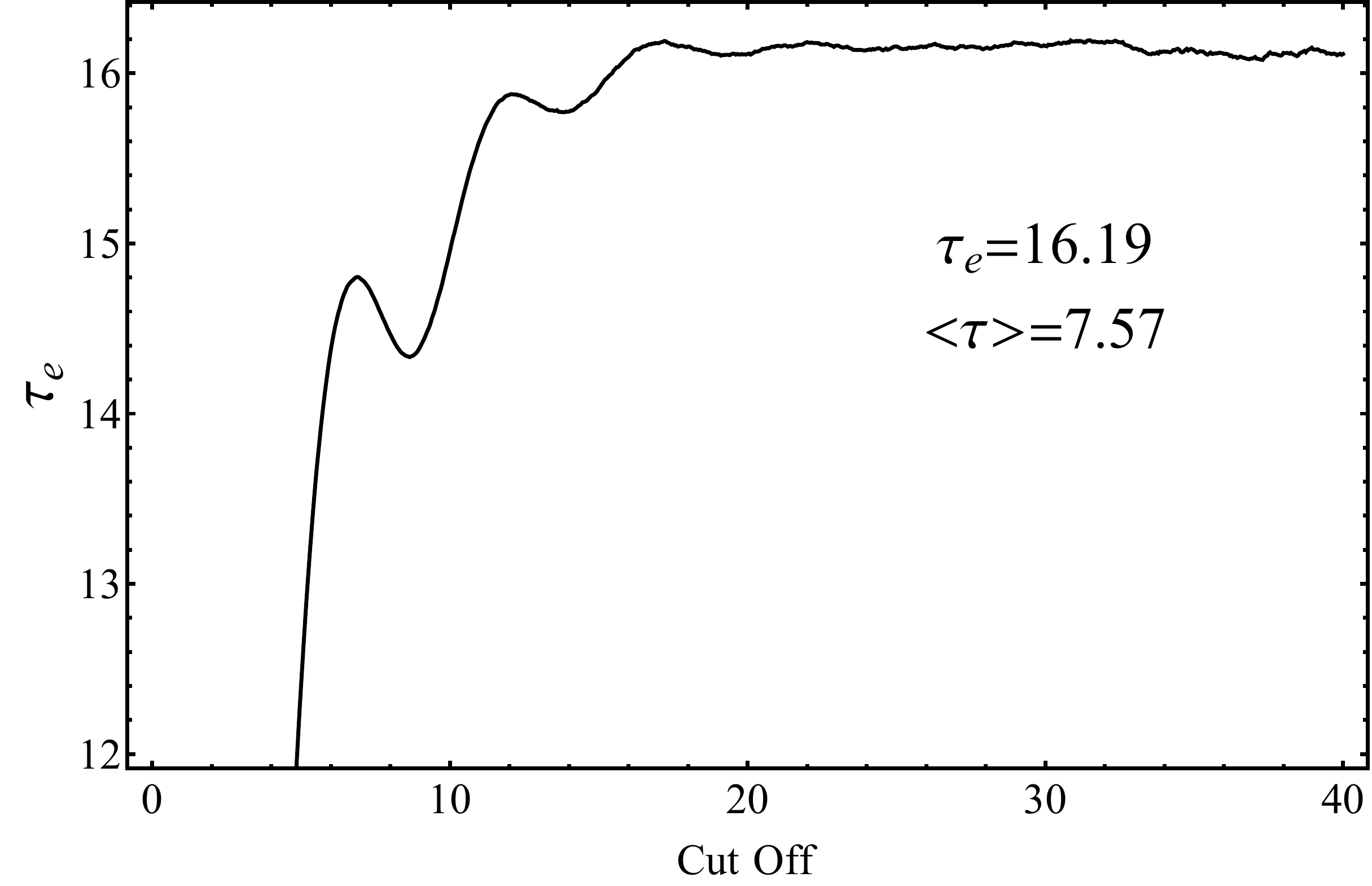}}
\caption{
Effect of the cut off on the estimate of the asymptotic escape time parameter $\tau_e$, the slope of Eq.(\ref{eq:distribution}) for long escapes.
The abscissa displays the cut off discussed in connection with Eq.(\ref{eq:fitting}).
}
\label{fig:cut}
\end{figure}

The outcome of the escape times sequence so far obtained can be processed to retrieve the average escape time. As mentioned before, we cannot use a straightforward sample mean estimator to ret rive the average escape time for the presence of deviations from the exponential distribution Eq.(\ref{eq:distribution}). 
To tackle this problem we have introduced a cut off in the escape time sequence. The sample mean is applied to the censored sequence, as shown in Fig. \ref{fig:cut}. 
In more details this amounts to adopt the following procedure:
\begin{enumerate}
\item Generate an escape time sequence;
\item Censor the obtained sequence below the cut off value $\tau_c$;
\item Apply a Maximum Likelihood Estimator of the parameter $\tau_e$ of Eq.(\ref{eq:fitting});
\item Repeat for different value of the cut off time $\tau_c$ until the estimate stabilizes within the selected approximation.
\end{enumerate}

We note that the result converges toward the asymptotic value for large enough cut off. An empirical rule of thumb to estimate the cut off is to take the first maximum of the escape times (see Fig. \ref{fig:distribution}a).

\section{Conclusion}

We consider a Fabry-Perot interferometer used to measure relevant properties of gravitational physics, in particular the detection of gravitational waves emitted by astrophysical sources. 
This is also a paramount problem of  fundamental physics measurements to demonstrate the presence of gravitational waves and, hopefully, to discriminate different gravitational theories among the many existing.
In this context a crucial role is played by the internal thermal noise of the end test masses. 
We have exploited the fact that the masses oscillations can drive the system across special points where abrupt reflectivity changes occur. 
Thus, the flight times between two reflectivity changes define escape  in the optomechanical potential (see Fig. \ref{fig:potential}) that carry information about the system and the presence of a coherent component.
To correctly collect the data, it is essential to define a suitable threshold that avoids transient (and spurious) oscillations around the separatrix, 
The analysis of the PDF of the escapes aims to extract the asymptotic behavior of the distribution (\ref{eq:distribution}). 
To do so, we have proposed a maximum likelihood estimator of the exponential parameter. 
This parameter has been compared with average escape time over the whole distribution.
and how to extract asymptotic exponential decay from censored data. 
We speculate that these techniques are of general interest for bistable systems \cite{Addesso15b}.


\section*{Acknowledgment}

The authors would like to thank Prof. I. M. Pinto for suggestions.

\end{document}